\documentclass[12pt,preprint]{aastex}

\def\um{\hbox{\,$\mu$m}}
\def\hr{\hbox{$^\mathrm{h}$}}
\def\m{\hbox{$^\mathrm{m}$}}
\def\s{\hbox{$^\mathrm{s}$}}
\def\msun{\hbox{\,M$_\odot$}}
\usepackage{color}
\definecolor{grass}{rgb}{0,0.5,0}
\definecolor{grey}{rgb}{0.5,0.5,0.5}

\slugcomment{}

\shorttitle{A Debris Disk around a 40\,Myr old M-type star}
\shortauthors{Teixeira et al.}

\begin{document}


\title{IRS Characterization of a Debris Disk around an M-type star in NGC\,2547}


\author{Paula S. Teixeira\altaffilmark{1,2,3,6}, 
        Charles J. Lada\altaffilmark{1},
	Kenneth Wood\altaffilmark{4},
	Thomas P. Robitaille\altaffilmark{4},
	Kevin L. Luhman\altaffilmark{5}
}
\altaffiltext{1}{Harvard-Smithsonian Center for Astrophysics, 60 Garden Street, Cambridge, MA 02138; \hbox{pteixeira@cfa.harvard.edu}, clada@cfa.harvard.edu}
\altaffiltext{2}{Departamento de F\'{\i}sica da Faculdade de Ci\^encias da \hbox{Universidade} de Lisboa, Ed. C8, Campo Grande, 1749-016, \hbox{Lisboa}, Portugal}
\altaffiltext{3}{Laborat\'orio Associado Instituto D. Luiz - SIM, Universidade de Lisboa, Campo Grande, 1749-016, \hbox{Lisboa}, Portugal}
\altaffiltext{4}{School of Physics and Astronomy, University of St. Andrews, North Haugh, St. Andrews KY16 9SS, UK; kw25@st-andrews.ac.uk, tr9@st-andrews.ac.uk}
\altaffiltext{5}{Department of Astronomy and Astrophysics, The Pennsylvania State University, University Park, PA 16802; kluhman@astro.psu.edu}
\altaffiltext{6}{Currently at the European Organisation for Astronomical Research in the Southern Hemisphere, Karl-Schwarzschild-Strasse 2, D-85748 Garching bei M\"unchen, Germany; pteixeir@eso.org.}

\begin{abstract}
We present 5 to 15\um\ \emph{Spitzer} Infrared Spectrograph (IRS) low resolution spectral data of a candidate debris disk around an M4.5 star identified as a likely member of the $\sim$40\,Myr old cluster NGC\,2547.  The IRS spectrum shows a silicate emission feature, indicating the presence of warm, small, (sub)micron-sized dust grains in the disk. Of the fifteen previously known candidate debris disks around M-type stars, the one we discuss in this paper is the first to have an observed mid-infrared spectrum and is also the first to have measured silicate emission.
We combined the IRS data with ancillary data (optical, $JHK_s$, and \emph{Spitzer} InfraRed Array Camera and 24\um\ data) to build the spectral energy distribution (SED) of the source. Monte Carlo radiation transfer modeling of the SED characterized the dust disk as being very flat ($h_{100}$=2\,AU) and extending inward within at least 0.13\,AU of the central star. 
Our analysis shows that the disk is collisionally dominated and is likely a debris disk.

\end{abstract}

\keywords{Galaxy: open clusters and associations: individual: NGC number: NGC 2547, infrared: stars, stars: circumstellar matter, stars: individual: 2MASS 08093547-4913033}

\section{Introduction}
\label{sec:intro}

The \emph{Spitzer} Space Telescope \citep{werner04} has provided the astronomical community with a wealth of information on sources with circumstellar disks, both previously known and recently discovered. The importance of having a more complete statistical sample of sources that have disks is tied in closely with our knowledge of the timescales for planet formation. For example, knowing when the inner region of the primordial or protoplanetary disk begins to clear itself of dust constrains the lower end of this timescale. Several ground-based observational near-infrared studies give evidence pointing to the clearing of the inner disk by 3 to 5\,Myr  \citep[e.g.][]{haisch01b}, yet the current paradigm is that planet formation occurs on timescales of 10\,Myr or longer \citep[e.g.][]{hillenbrand05}. 
The primordial disk is essentially an accretion disk (formed as the parental molecular cloud collapsed into a star) that is still feeding the newly formed star with material. Mainly, magnetospheric accretion is responsible for the clearing of most of the gas and dust in the disk \citep[e.g.][]{muzerolle01}, however, planet formation also helps with the clearing of the gaseous protoplanetary disk by the the formation of planetesimals that sweep up much of the disk's material. The formation of larger planetary bodies occurs via collisions of smaller planetesimals \citep[lunar-mass oligarchs, see][and references therein]{kenyon06,chambers01} and this process gene\-rates a secondary dusty disk, mostly devoid of gas, known as a debris disk, i.e., the debris disk consists mainly of dust that is produced in the collisions between the newly formed planetesimals. Since there need to be colliding planetesimals present to replenish the dust in the disk, detecting a debris disk is a strong signature of an emerging planetary system. 

To constrain the duration of the primordial accretion disk phase and the transition from primordial to debris disks it is necessary to analyze stars of intermediate age (between 20 and 40\,Myr). 
The NGC\,2547 stellar cluster presents itself as an ideal candidate to study the evolution of circumstellar disks because of its relatively close distance, 361$^{+19}_{-8}$\,pc, and age 38.5$^{+3.5}_{-6.5}$\,Myr \citep{naylor06}.
To investigate the disk population of this cluster, \citet{young04} acquired \emph{Spitzer} data (Program ID 58, P.I. George Rieke) of the cluster NGC\,2547, consisting of imaging in all four InfraRed Array Camera (IRAC) \citep{fazio04} bands and all three Multiband Imaging Photometer for \emph{Spitzer} (MIPS) \citep{rieke04} bands. The preliminary results were reported in \citet{young04}, where they found that a very small fraction ($\sim$\,5\%) of cluster members show evidence for levels of infrared excess emission likely due to primordial disks. 
The other significant result from \citet{young04} was the discovery of a candidate late-type star (2MASS\,08093547-4913033) that has a considerable 24\,$\mu$m excess. It presents no detectable excess emission shortward of 8\um, while the 24 um excess is reduced compared to CTTS, suggesting it may possess a debris disk with an inner hole. 
This star is of interest because of the rarity of observations of debris disks around late-type stars \citep[e.g.][]{lestrade06,siegler07,gautier07,forbrich08}, especially for NGC\,2547's age. 
Observing a large excess emission, such as that detected towards this particular star, could be indicative of recent planet formation. Numerical calculations carried out by \citet{kenyon02} showed that formation of planetary bodies that are 1000\,km in size or larger produces copious amounts of dust, creating greater scattering of light from the central star and producing excess emission.

The aforementioned IRAC and MIPS data were re-analyzed and recently published by \citet{gorlova07}; from their membership list they found that $<$1\% of sources show IRAC excesses, while 30-45\% of B-F members present 24\um\ excess emission. \citet{gorlova07} also discovered an additional M dwarf with excess emission indicative of the presence of a debris disk.
To better probe the fraction of debris disks among lower mass stars \citet{forbrich08} acquired an extremely deep MIPS dataset (Program ID 21127, P.I. Charles J. Lada) and discovered nine additional candidate M-dwarf debris disks. Although M-type stars are the most numerous of any type, their debris disk statistics are extremely scarce due to their faintness. Acquiring more observations of debris disks around M stars is therfore very important to improve our understanding of how these disks evolve and planet formation occurs.

In this paper we report our findings regarding the debris disk around the aforementioned member of NGC\,2547, 2MASS 08093547-4913033, found by \citet{young04} to possess considerable excess emission at 24\um. This source will be referred to henceforth as source \#\,23, following \citet{forbrich08}'s nomenclature. To determine the source's spectral type, we acquired optical spectroscopy. We obtained \emph{Spitzer} Infrared Spectrograph (IRS) observations to study this source in more detail, in particular, to determine whether the disk has silicate emission and to help constrain its spectral energy distribution (SED) modeling. The shape of silicate features gives insight into dust grain size and possibly also its crystallinity if the signal-to-noise ratio is high enough. 

The organization of the paper is as follows: \S\,\ref{sec:obs} summarizes the acquisition and reduction of the optical and IRS spectroscopic data, and \S\,\ref{sec:res} presents the results obtained. We discuss the implications of our results in \S\,\ref{sec:dis}, and finally, we list our conclusions in \S\,\ref{sec:concl}.

\section{Observations and data reduction}
\label{sec:obs}
\subsection{\emph{Spitzer} InfraRed Spectrograph data}

\begin{figure}
\centering
\includegraphics[width=0.45\textwidth]{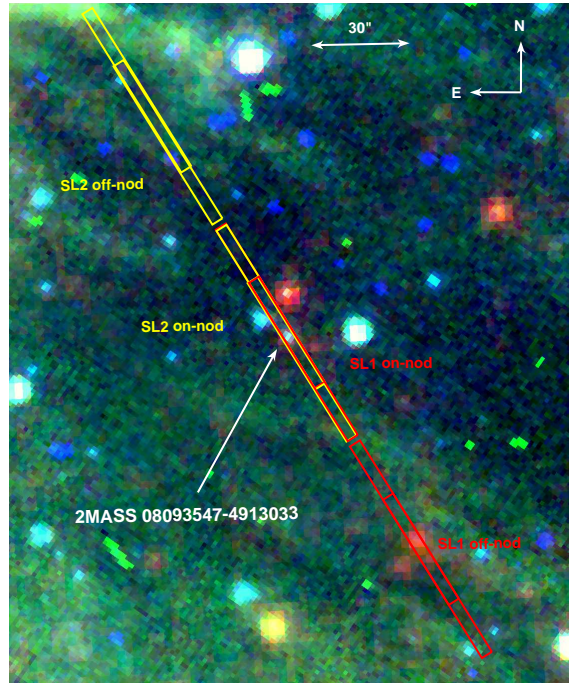}
\caption{Color composite image where red, green, and blue correspond to MIPS 24\um, IRAC 8\um\ and 5.8\um\ data, respectively. The IRS SL AORs used are overplotted in red (1st order) and yellow (2nd order). The target source (2MASS\,08093547-4913033 or source \#\,23) is indicated by an arrow.}
\label{fig:aor}
\end{figure}

The spectroscopic data of source \#\,23 was acquired on the 19th of June of 2006 (IRS campaign 32), through GO proposal Program ID\,20750 (P.I. Charles J. Lada). The data are comprised of low resolution (R\,$\sim$\,60-127) observations taken using the \emph{Spitzer} IRS \citep{houck04a} Short-Low (SL) module, with a total on-source integration time of 3.2 hours.
 We built four Astronomical Observation Requests (AORs), each with 2 cycles of 240\,s of ramp duration for the second order of SL (SL2: 5.2-8.7\um) and 10 cycles of 240\,s of ramp duration for the first order of SL (SL1: 7.4-14.5\um); the AOR keys are  15096064, 15096320, 15096570, and 15096832. We used a pointing control reference sensors (PCRS) peak-up target located at ($\alpha,\delta$)(J2000)=(08\hr09\m59.22\s, -49\degr16\arcmin11.0\arcsec), with m$_\mathrm{V}=8.14$ magnitudes. Figure \ref{fig:aor} shows the orientation of the AORs with respect to the background in NGC\,2547 during the observations.\\

\begin{figure}
\centering
\includegraphics[width=0.45\textwidth,angle=90]{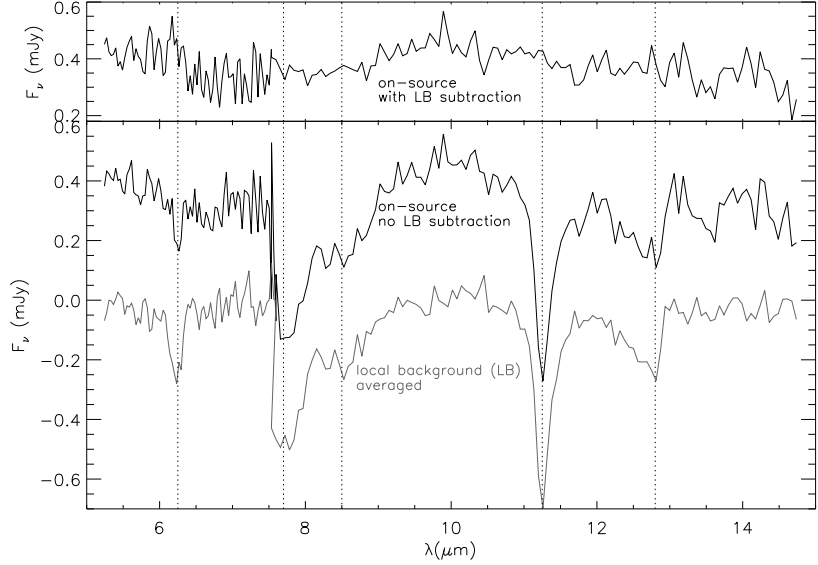}
\caption{Comparison of three spectra: the bottom panel shows the averaged local background spectrum (gray) and the on-source spectrum with no local background subtraction (black), while the top panel shows the on-source with local background subtraction spectrum. The PAH bands (6.25, 7.7, 8.5, 11.25, and 12.8\um) are marked by dotted vertical lines.}
\label{fig:spectra}
\end{figure}

We used the CUbe Builder for IRS Spectra Maps (CUBISM) software \citep{smith07} to reduce and extract the spectrum of the source from the four AORs, for each order. 
The reduction began with the subtraction of the off nod position slit from the on-source position slit for each order; this step was meant to (i) eliminate instrumental effects such as leftover background gradients from the pipeline flat-fielding, (ii) subtract out the zodiacal emission, and (iii) remove rogue pixels (created mostly by cosmic rays). Additional bad pixels were identified and removed. Three spectra were extracted for each order, namely, on-source spectrum, and two local background spectra on either side of the source. The SMART software \citep{higdon04} was then used to merge the first and second orders, to construct an averaged local background spectrum, and also used to build the final on-source spectrum with local background subtraction. Figure \ref{fig:spectra} shows the average local background used (gray) and on-source spectrum with and without local background subtraction. Broad polycyclic aromatic hydrocarbons (PAHs) bands are present as absorption features in the averaged local background spectrum and on-source with no local background subtraction spectrum - these are in fact emission bands present in the off-nod observations (Fi\-gure \ref{fig:aor} shows faint diffuse emission in the off-nod slits). As mentioned above, the off-nod observations were subtracted from the on-nod observations, turning the PAH emission bands into ``absorption'' bands. The local background subtraction adds back in the PAH flux removed by the on-off slit subtraction (see Figure \ref{fig:spectra}). From 7.5\um\ to 14.8\um\ the spectrum has a signal-to-noise ratio (SNR) of
33, and from 5.2\um\ to 7.5\um\ the SNR is 22.

\subsection{Optical spectroscopy at Magellan}

\begin{figure}
\centering
\includegraphics[width=0.5\textwidth]{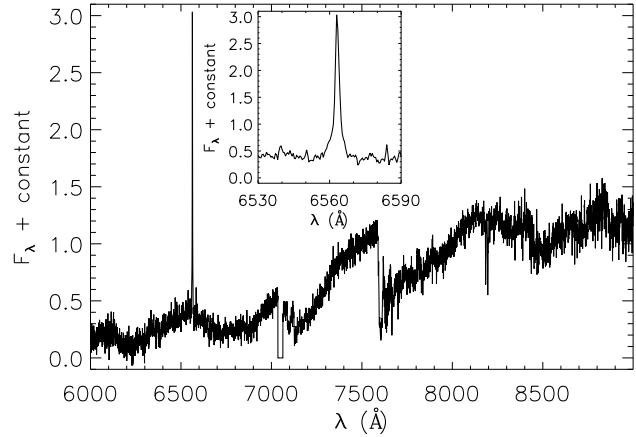}
\caption{Optical spectrum of the source \#\,23. The inset is a zoom-in on the H$\alpha$ emission line. Based on a comparison of these data to spectra of field dwarf standards, we classify this star as M4.5. The spectrum is normalized at 7500\,\AA.}
\label{fig:optical_spectra}
\end{figure}

Through Spitzer imaging of NGC2547, \citet{young04} detected excess emission at 24\um\ from the candidate low-mass star \#\,23. We have used an optical spectrum of this source to measure its spectral type and to constrain the presence of accretion through the strength of the H$\alpha$ emission line. The spectrum was obtained on 2004 April 26 during multi-slit spectroscopic observations of candidate members of NGC\,2547 using the Inamori Magellan Areal Camera and Spectrograph (IMACS) on the Magellan\,I telescope at Las Campanas Observatory. The spectrograph was operated with the 600~\,l\,mm$^{-1}$ grism, OG570 blocking filter, and slit widths of $0.9\arcsec$, which produced a spectral resolution of 2.0~\AA. The exposure time was 60\,s. After bias subtraction and flat-fielding, the spectrum of the source was extracted and calibrated in wavelength with arc lamp data. We corrected for the sensitivity functions of the detectors through observations of a spectrophotometric standard star. The optical spectrum of the star \#\,23 is presented in Figure\,\ref{fig:optical_spectra}.

\section{Results}
\label{sec:res}

Through a comparison to standard field dwarfs, we have measured a spectral type of M4.5 and an equivalent width $W_{\lambda}=17$~\AA\ for H$\alpha$ from the optical spectrum. The spectral type we determine agrees with that obtained by \citet{jeffries05}.

\begin{figure}
\centering
\includegraphics[width=0.5\textwidth,angle=90]{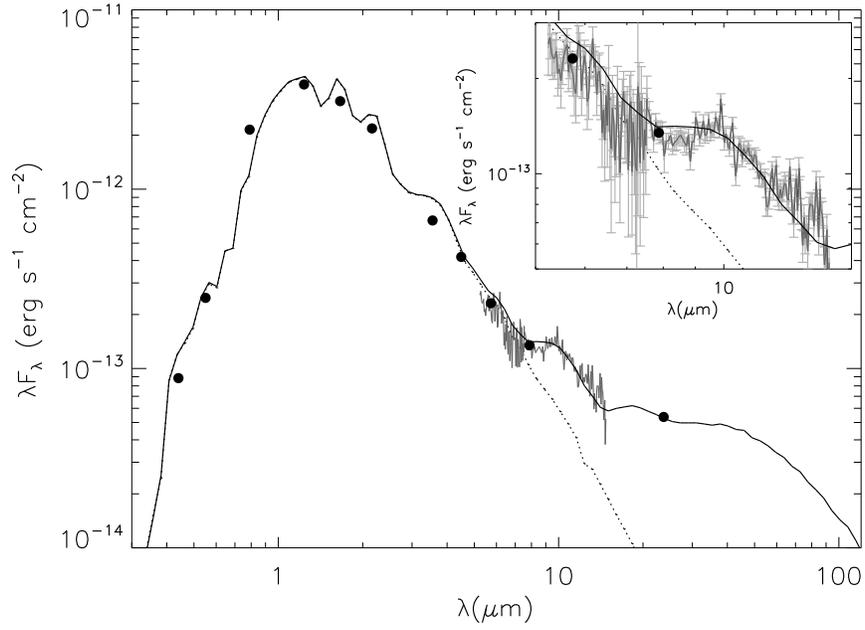}
\caption{Spectral energy distribution of source \#\,23. The photometric data points are marked as filled circles and are comprised of BVI, 2MASS, IRAC, MIPS 24\um; the IRS SL data is represented by the solid gray line. The black curve corresponds to the best fit model, whose parameters are given in Table \ref{tab:sed}, and the dotted curve represents the photosphere. The inset shows a zoom-in on the IRS spectrum and the model fit (the spectrum errorbars are represented in lighter gray).}
\label{fig:sed}
\end{figure}

The on-source, local background subtracted, IRS spectrum of Figure \ref{fig:spectra} was combined with existing multiband wavelength photometry, from the optical to MIPS 24\um\ \citep{jeffries04,young04,forbrich08} to build an SED, shown in Figure \ref{fig:sed}. The final spectrum shows silicate emission, however the variable background and the faintness of the source makes it very difficult to draw strong conclusions about the shape of this feature. 

We model the SED using Monte Carlo radiation transfer codes developed for modeling and interpreting spectra and images of dusty accretion disks and envelopes \citep[see][for a description of the codes]{whitney03a,whitney03b}. The codes include multiple anisotropic scattering and use the \citet{bjorkman01} algorithm for computing radiative equilibrium dust temperatures and thermal spectra.

The presence of a silicate feature in the SED indicates that the disk contains small grains \citep{weingartner01}, so we adopt the \citet{kim94} interstellar medium (ISM) dust model for the opacity and scattering properties \citep[see][for a graphical representation of the wavelength dependence of this model]{whitney03a,whitney03b}. The grain size distribution of \citet{kim94} model follows a power law (slope $\sim$ -3) between 0.02\um\ and 0.2\um, and then falls off exponentially.

 For the object under study we adopt an axisymmetric flared disk density structure and assume the disk is heated by starlight. The density structure is parameterized by
\begin{equation}
\rho (r,z)= \rho_0 \exp{\left[- \frac{1}{2}\left(\frac{z}{ h}\right)^2\right]} \left({R\over{r}}\right)^{\alpha},
\end{equation}
where $r$ is the cylindrical radius, $z$ is the vertical scale, $\alpha$ is the radial density exponent, and $\rho_0$ is the normalization constant. The disk scale-height is given by $h(r)=h_{100} (r/{\rm 100\; AU})^\beta$, where $h_{100}$ is the disk scale-height at a radius of 100\,AU and $\beta=\alpha -1$ is the flaring power. The disk extends from an inner radius, $R_{\rm min}$ to an outer radius $R_d$. 

The lack of longer wavelength data makes it difficult to constrain disk parameters, so we fix some parameters based on previous models of irradiated disks. We adopt a surface density of the form $\Sigma(r)\sim r^{-1}$ which is typical of models of irradiated disks and fix $\beta=1.25$ and $\alpha = 2.25$ \citep[e.g.][]{d'alessio98}. The SED is not very sensitive to the disk outer radius \citep[e.g., see models in][]{scholz06}, so we set $R_d = 300$\,AU.

For the central star we use a temperature of $T_\star = 3089$, a radius of $R_\star$ = 0.47\,$M_\odot$, a luminosity of $L_\star = 0.017 L_\odot$, and the spectrum is for a 3000\,K NextGen model atmosphere \citep{hauschildt99a}.
With these star, disk, and dust parameters fixed, we constructed a small grid of SED models varying $R_{\rm min}$, $h_{100}$ and total dust disk mass, $M_d$. The SED is somewhat degenerate in $h_{100}$ and $M_d$ and we find a good model fit using $h_{100}=2$\,AU, $M_d = 2.7 (R_d/300\,AU) M_{lunar}$, and disk luminosity of $L_{disk}=0.035 L_\star$. The degeneracies in these models may be broken with longer wavelength data which is sensitive to the disk mass. We see that the silicate feature is fit rather well using the adopted ISM grain model. However, this does not rule out the possibility that grain growth has occurred in the disk and the silicate feature is produced by a population of small grains in the upper disk layers with larger grains settled towards the midplane \citep[e.g.][]{scholz07}. 

The best fit model is shown in Figure \ref{fig:sed} and the corresponding parameters are summarized in Table \ref{tab:sed}.
What we can say from the modeling is that the disk is considerably lower mass and flatter than T~Tauri disks. Our models also indicate that if there is an inner disk hole it is not very large, $R_{\rm min}\leq 0.13$\,AU, smaller than that found in other debris disks, e.g. AU Mic, typically many AU. However, the nominal model suggests that the hole is larger than the dust sublimation radius.

\section{Discussion}
\label{sec:dis}

\subsection{The nature of the disk}
\label{subsec:timescales}

Although  the results derived from SED modeling described above are consistent with a homologously evolved protoplanetary disk \citep{wood02b}, the disk appears to be devoid of gas and non-accreting: the equivalent width of the H$\alpha$ emission line is 17\AA, which for the spectral type in question could originate in the stellar chromosphere \citep{white03}.

To better ascertain the nature of this disk, we first consider the relevant timescales for debris disk clearing and apply them for our source.
There are four possible mechanisms for the removal of dust grains from within a debris disk \citep[e.g.][]{draine79,kruegel03,grigorieva07b}: (i) sublimation, (ii) stellar radiation pressure, (iii) stellar wind, and (iv) Poynting-Roberston drag.
The sublimation temperatures for dust grains depend on their composition: amorphous carbon grains sublimate at 2000\,K, whereas silicate grains sublimate at 1500\,K \citep[e.g.][]{preibisch93}. 
These temperatures are only reached at very close proximity to the star\footnote{using the radial dust temperature profile given by equation 28 from \citet{kamp00}, we determine that the sublimation radius of silicate grains of average size of 0.2\um\ around source \#23 is 0.0045\,AU or 2.0\,R$_\star$.}; sublimation may be an important factor in the production of the very innermost regions of a hole, but it is not a relevant grain removal mechanism for the debris disk being discussed. The radiation pressure is also an insignificant  grain removal mechanism for debris disks around M mass stars \citep[we refer the reader to ][ for a detailed discussion on this matter]{plavchan05}. Regarding stellar wind drag, \citet{plavchan05} propose it may be a dominant grain removal process, but only for very young and active M dwarfs with extremely high mass loss rates ($\sim$\,10$^3$ times higher than solar or 10$^{-11}$\msun yr$^{-1}$), which is not the case for source \#\,23. Using the mass and radius of Table \ref{tab:sed} we calculated the escape velocity for this source to be 299\,km\,s$^{-1}$.
A Gaussian fit to the H$\alpha$ emission line profile gives a full with at half maximum of 114\,km\,s$^{-1}$, so it seems winds are not very energetic for this source.

The P-R drag timescale, i.e., the time is takes for a dust grain to spiral inwards into the star, is given by \citep{kruegel03}:

\begin{equation}
\tau_{P-R}=\frac{4 \pi \rho_{gr} a c^2 r^2}{3 L_\star}
\label{eq:pr1}
\end{equation}

\noindent where $a$ and $\rho_{gr}$ are the radius and density of the dust grain, respectively, and $c$ is the speed of light in the vacuum. 

\emph{Spitzer} IRS spectra of the Deep Impact ejecta of comet 9P/Temple\,1 revealed a dust mixture composed of amorphous and crystalline silicates, amorphous carbon, water ice, and other components, with a mean density of the dust grains for this composition of 2.96\,g\,cm$^{-3}$ \citep{lisse08}\footnote{the densities of amorphous carbon, graphite, olivine, forsterite, and enstatite are 1.81, 2.16, 3.32, 3.27, and 3.20\,g\,cm$^{-3}$, respectively.}. A similar dust composition was found in the debris disk of HD\,113766\,A by \citet{lisse08}, who observed the source using 5-35\,$\mu$m \emph{Spitzer} IRS data. We assume the dust in the debris disk of source \#\,23 is similar in composition to that of HD\,113766\,A or comet 9P/Temple\,1 and we henceforth use the same grain density for our calculations.
Equation \ref{eq:pr1} is then re-written solely as a function of $a$ and $r$:

\begin{equation}
\tau_{P-R}=0.12\left(\frac{a}{\mu m}\right)\left(\frac{r}{AU}\right)^2\ \ Myr.
\label{eq:pr2}
\end{equation}

\noindent A 0.2\um-sized dust grain at 1\,AU would then take 2.4\,$\times$\,10$^{4}$ years to spiral onto our source. On the other hand, a 0.2\um-sized grain located at 41\,AU would take $\sim$\,40\,Myr to spiral onto the star, so small grains located beyond this radius could be leftover material from the primordial accretion disk if no other dust removal or reprocessing mechanism is at play. 

Yet there is one more important physical process to be considered: grain-grain collisions. These collisions either shatter the dust grains into smaller particles (high collision velocities), or help clear the disk by triggering grain growth through agglomeration (low collision velocities); grain-grain collisions therefore re-distribute the grain size population. Grain-grain collisions will also act to reduce the vertical scale height of the disk \citep[see e.g. ][]{besla07}. The best-fit SED model for our source does indeed predict a ``razor-thin'' or very flat disk, with a scale height of 2\,AU at 100\,AU (see \S\,\ref{sec:res}, Table \ref{tab:sed}).  The grain-grain collision timescale, $\tau_{coll}$, is given by \citep{backman93}:

\begin{equation}
\tau_{coll} \sim \frac{\tau_{orb}}{\sigma}
\label{eq:tcol_0}
\end{equation}

\noindent where $\tau_{orb}$ is the orbital period and $\sigma$ is the dimensionless fractional surface density; the latter can be approximated by $L_{disk}/L_\star$ \citep{moro-martin07a,moro-martin07c} and Equation \ref{eq:tcol_0} is written as:

\begin{equation}
\tau_{coll} \sim 0.125 \left(\frac{r}{AU}\right)^{3/2}\left(\frac{M_\odot}{M_\star}\right)^{1/2}\frac{L_\star}{L_{disk}}\ \ yr
\label{eq:tcol}
\end{equation}

\noindent where $r$ is the distance of the grain from the star and $M_\star$ is the stellar mass. 
Combining Equations \ref{eq:pr2} and \ref{eq:tcol}, using the disk luminosity and stellar mass cited in Table \ref{tab:sed}, and a grain size of 0.2\um, one finds that the collisional timescale is shorter than the P-R drag timescale for a disk radius greater than 2.0\,$\times$\,10$^{-7}$\,AU. In other words, the disk is everywhere collisionally dominated. This implies that the disk is not a remnant protoplanetary disk but a bona fide debris disk since the dust grains have been reprocessed and are ``second-generation''.

For disks with lower relative luminosities, the inner regions may be dominated by P-R drag, however, the outer regions are always collision-dominated. For example, a disk with $L_{disk}=3.5\,\times\,10^{-5}L_\star$ with average grain sizes of 0.2\um, would have an inner hole of (at least) 0.2\,AU radius because in that inner region the P-R drag timescale is shorter than the grain-grain collision timescale, i.e., the dust grains would spiral in towards the star at a faster rate compared to the rate at which they are produced by grain-grain collisions.

To date, there are fifteen known debris disks around M stars, namely, AU\,Mic (GJ\,803, HD\,197481), M0.5 Ve GJ\,182, M1 TWA\,7, M0.5 GJ\,842.2, and eleven sources in NGC\,2547 \citep{forbrich08}. AU\,Mic is known to have an inner hole of 17\,AU and \hbox{$L_{disk}/L_\star=6\,\times\,10^{-4}$} \citep{liu04a}. Equations \ref{eq:pr2} and \ref{eq:tcol} could be used to explain an inner hole of $\sim$0.5\,AU if the grain sizes were 0.01\um, so in AU Mic's  particular system another mechanism may be clearing the dust \citep[such as substellar or planetary companion, or stellar wind drag, ][respectively]{liu04a,plavchan05}. As for GJ\,182, it has 24\um\ excess and if an inner hole exists it has a radius smaller than 1\,AU according to \citet{liu04a}. TWA\,7 also has 24\um\ excess emission \citep{low05} although it has not yet been possible to determine if the disk has an inner hole \citep{matthews07}. \citet{lestrade06} determined that the disk associated with GJ\,842.2 has its peak emission at 140\,AU although there is tentative 25\um\ emission indicative of warm dust.

The measured disk masses of previously known debris are reported as 28\,M$_{lunar}$ \citep[GJ\,842.2 ][]{lestrade06}, 18\,M$_{lunar}$ \citep[TWA\,7][]{matthews07}, 2.11\,M$_{lunar}$ \citep[GJ\,182][]{liu04a}, and 0.89\,M$_{lunar}$ \citep[AU\,Mic][]{liu04a}. The fractional disk luminosities, $L_{disk}/L_\star$ vary between 2$\times$10$^{-3}$ \citep[TWA\,7 ][]{low05} and 6$\times$10$^{-4}$ \citep[Au\,Mic][]{liu04a}. Regarding disk sizes, \citet{lestrade06} determined a radius of 300\,AU for GJ\,842.2 and AU\,Mic's disk has a radius of 210\,AU. Finally, there is a wide spread in the ages for these sources, from 200\,Myr \citep[GJ842.2][]{lestrade06} to 12\,Myr \citep[AU\,Mic][]{strubbe06}.
Although the disk around source \#23 has an order of magnitude higher fractional luminosity than these other debris disks, it appears to be produced by a comparable amount of dust.

The disk scale height we determined from the SED modeling of the debris disk, 2\,AU at 100\,AU, is similar to AU\,Mic disk's scale height, 2.5\,AU. These scale heights are much smaller than those of T\,Tauri protoplanetary disks. For comparison, \citet{wood02a} modeled the SED of the T Tauri star HH\,30\,IRS, using the same Monte Carlo radiation transfer code as used in the work presented in this paper, and found that the disk scale height was 17\,AU at a disk radius of 100\,AU radius. Another T Tauri star whose SED was modeled using the same aforementioned code is GM\,Aurigae, that was found to have a scale height of 8\,AU at 100\,AU radius.

\subsection{On the silicate feature and the disk mass}
\label{subsec:mass}

As mentioned above in \S\,\ref{sec:res}, the silicate feature in emission indicates that there are small grains in the debris disk.
Unfortunately, due to the PAH contamination and the faintness of the source, we cannot draw any conclusions regarding silicate crystallization. Although we cannot draw more quantitative conclusions from our IRS observations, to our knowledge these observations correspond to the \emph{first silicate emission detection from a debris disk around and M-type star}.

To estimate the temperature of silicate grains we assume their temperature is that of a blackbody with an SED peaking at 10\um, i.e., 290\,K. The temperature profile obtained from the SED modeling (\S\,\ref{sec:res}) indicates that the disk reaches that temperature at a radius of $\sim$0.15\,AU (although the silicate emission could also be arising from the upper layers of the disk). The snowline, defined as the location where the temperature corresponds to that of water ice sublimation \citep[153\,K,][]{hayashi81}, for this disk  would be located at $\sim$0.18\,AU. The habitable zone, $a_\mathrm{HZ}$, defined as the region in the disk where a planet can sustain liquid water at its surface, is centered roughly on a radius of 0.13\,AU for our source \citep[using the relation $a_\mathrm{HZ}=\sqrt{L_\star/L_\odot}$ from ][]{scalo07}. To date there are seven known planet-host M dwarf stars, and these have orbits ranging from 0.02 to 2.3\,AU \citep[][ and references therein]{endl08}. So, planet formation can indeed occurr within the habitable zone or within the snowline for our source. In other words, there could be terrestrial planet formation occuring within the habitable zone, however, more data is needed to study this hypothesis in more detail.

Among the M dwarf debris disk candidates in NGC\,2547, source \#23 is the one that presents the biggest excess emission at 24\um\ \citep{forbrich08}. Such a large excess, indicative of a relatively luminous disk, may be the result of a stochastic collision of planetary bodies within the disk \citep[e.g.][]{kokubo06} that triggered a collisional cascade \citep{kenyon02}. Such a catastrophic event happened in our own solar system, and it led to the formation of the Moon \citep{kleine02}.

The dust disk mass determined from the SED fitting, 1\,$\times$\,10$^{-7}$\,M$_\odot$ (or 2.7\,M$_{lunar}$),  is dependent of the distribution of the grain sizes. The fact that we detect silicate emission is an indication that small ISM-like dust grains are indeed present in the disk. Unfortunately, no longer wavelength data is available and we are unable to further constrain the grain sizes used in our model. We can however assume that the grain sizes are smaller than those for debris disks around earlier type stars since radiation pressure is not a significant grain removal process for M stars \citep[e.g., the radiative blowout grain size for an A0-type star is 7.7\um,][]{besla07}. Furthermore, in \S\,\ref{subsec:timescales} we showed that the grain-grain collisional timescale for the system is shorter than the P-R drag timescales, and consequently  many small grains, possibly submicron-sized,  are produced and remain in the disk. We also showed that there is no effective grain removal mechanism for small grains in the disk, for the exception of grain growth. We believe our mass estimate is therefore a good approximation of the true disk mass for the SED fit we obtain.

\section{Summary}
\label{sec:concl}

We have acquired \emph{Spitzer} low resolution spectral data from 5 to 14\um\ of a M4.5 star, 2MASS\,08093547-4913033 or source \#\,23, that is a candidate member of NGC\,2547, a $\sim$40\,Myr old cluster. Combined with previously published optical, near-infrared, IRAC, and MIPS data, we built the SED of the source and modeled several stellar and disk parameters. We briefly summarize the main results of the paper as follows:
\begin{enumerate}
\item We confirm excess emission longwards of 8\um\ for the system and report the first detection of silicate emission from a debris disk around an M-type star;
\item The silicate emission indicates that the dust grains seem to be small: micron or sub\-micron-sized;
\item SED modeling describes the debris disk as being very flat (scale height of 2\,AU at 100\,AU radius), and as having a mass of $\sim$ 2.7 lunar assuming a disk radius of 300\,AU. The SED model suggests that the disk extends to within 0.13\,AU of the stellar surface, outside the dust sublimation radius.

\item Comparison of the various timescales involved in removing and producing dust leads us to conclude that the disk is dominated by grain-grain collisions throughout its entire extent;
\item A stochastic collision of planetary bodies could be responsible for the higher disk luminosity of this source, and formation of terrestrial planets within the habitable zone is entirely possible for this system.
\item We confirm the spectral type of M4.5 for the source and find that its H$\alpha$ emission is consistent with chromospheric activity; the H$\alpha$ emission line is also narrow enough to rule out strong stellar winds.
\end{enumerate}

\acknowledgments

We are very grateful to an anonymous referee for a careful review of the paper. We are also indebted to James Muzerolle for building the IRS AORs used, and last but not least, we thank Jan Forbrich and Achim Tappe for many useful discussions.
P.~S.~T. acknowledges support from the scholarship SFRH/BD/13984/2003 awarded by the Funda\c{c}\~ao para a Ci\^encia e Tecnologia (Portugal).
K.~L. was supported by grant AST-0544588 from the National Science Foundation.
This work is based in part on observations made with the \emph{Spitzer} Space Telescope, which is operated by the Jet Propulsion Laboratory, California Institute of Technology under a contract with NASA. Support for this work was provided by NASA.
The IRS was a collaborative venture between Cornell University and Ball Aerospace Corporation funded by NASA through the Jet Propulsion Laboratory and Ames Research Center. 
SMART was developed by the IRS Team at Cornell University and is available through the Spitzer Science Center at Caltech.

{\it Facilities:} \facility{\emph{Spitzer}(IRS)}.

\clearpage

\clearpage

\begin{deluxetable}{llc}
\tablecolumns{3}
\tablewidth{0pt}
\tablecaption{Stellar and disk parameters for source \#\,23.\label{tab:sed}}
\tablehead{
\colhead{Parameter}       & \colhead{Symbol (units)} & \colhead{Value}
}
\startdata
adopted distance          & $D$ (pc)                 & 361                            \\
stellar temperature       & $T_\star$ (K)            & 3089                           \\
stellar radius            & $R_\star$ (R$_\odot$)    & 0.47                           \\
stellar luminosity        & $L_\star$ (L$_\odot$)    & 0.017                          \\
stellar mass              & $M_\star$ (M$_\odot$)    & 0.11                           \\
adopted disk outer radius & $R_{disk}$ (AU)          & 300                            \\
disk mass                 & $M_{disk}$ (\msun)\tablenotemark{\dagger}       & 1$\times$10$^{-7}$\tablenotemark{\dagger}             \\
disk inner radius         & $R_{hole}$  (AU)         & $\leq$ 0.13                    \\
scale height at 100\,AU   & $h_{100}$ (AU)           & 2                              \\
disk luminosity           & $L_{disk}$ (L$_\star$)   & 0.035                          \\
\enddata
\tablenotetext{\dagger}{for a disk radius of 300\,AU.}
\end{deluxetable}
\clearpage 


\end{document}